\documentclass[a4paper, 11pt]{article}
\pdfoutput=1

\usepackage[margin=2.5cm]{geometry}
\usepackage{latexsym}
\usepackage{graphics}
\usepackage{epsfig}

\def\be{\begin{equation}}
\def\ee{\end{equation}}

\def\bea{\begin{eqnarray}}
\def\eea{\end{eqnarray}}

\begin{document}

\title{Graphene and Black Holes: \\ novel materials to reach the unreachable}
\author{Alfredo Iorio\thanks{E-mail: alfredo.iorio@mff.cuni.cz} \\
Faculty of Mathematics and Physics \\ Charles University in Prague \\ V Hole\v{s}ovi\v{c}k\'ach 2, 18000 Prague 8, Czech Republic
\\
alfredo.iorio@mff.cuni.cz}

\date{\today}

\maketitle

Physics is an experimental science. Nonetheless, in the last decades, it proved very difficult (if not impossible) to reconcile theoretical investigations of the fundamental laws of Nature with the necessary experimental tests. The divarication between theory and experiments is the central problem of contemporary fundamental physics that, to date, is still unable to furnish a consistent quantum theory of gravity, or to obtain experimental evidences of milestones theories like Supersymmetry (see, e.g., the whole Issue \cite{thooftstring} on the status of String Theory).

A widespread view of this problem is that experimental observations of these type of phenomena can only be achieved at energies out of the reach of our laboratories (e.g., the Planck energy $10^{19}$GeV vs reachable energies in our laboratories of $10^3$GeV). In our view, due to unprecedented behaviors of certain novel materials, nowadays indirect tests should be considered as a viable alternative to direct observations.

This field of research is not a novelty. It usually goes under the generic name of ``analogue gravity'' \cite{volovik1}, although, perhaps, to call it ``bottom-up'' approach does it more justice: analogue experimental settings on the bottom, fundamental theories of Nature on the top. Nonetheless, for various reasons, it has been seen as little more than a curiosity. An amusing and mysterious series of coincidences, that never (up to our knowledge) were taken as tests of those aspects of the fundamental theories that they are reproducing. Neither are taken as experimental tests of the fundamental theories, the mysterious and amusing coincidences of the ``top-bottom'' approach of the AdS/CFT correspondence (where theoretical constructions of the fundamental world,  are used to describe experimental results at our energy scale \cite{maldacena}).

As well known, with graphene \cite{geimnovoselovFIRST} we have a quantum relativistic-like Dirac massless field available on a nearly perfectly 2-dimensional sheet of carbon atoms (see \cite{pacoreview2009} for a review). Recent work shows the emergence of gravity-like phenomena on graphene \cite{iorio}.  More precisely, the Hawking effect can take place on graphene membranes shaped as Beltrami pseudospheres of suitably large size, hence even (2+1)-dimensional black-hole scenarios \cite{btz} are in sight. The Hawking effect here manifests itself through a finite temperature electronic local density of states. For reviews see \cite{reviewslectures}.

The predictions of the Hawking effect on graphene, are based on the possibility to obtain very specific shapes, e.g., the Beltrami pseudosphere, that should recreate, for the pseudoparticles of graphene, conditions related to those of a spacetime with an horizon. What we are first trying \cite{ComingTrumpet} is to obtain a clear picture of what happens to $N$ classical particles, interacting via a simple potential, e.g., a Lennard-Jones potential, and constrained on the Beltrami. This will furnish important pieces of information on the actual structure of the membrane. In fact, is well known from similar work with the sphere (that goes under the name of ``generalized Thomson problem'', see, e.g., \cite{bowick}) that defects will form more and more, and their spatial arrangements are highly non trivial, and follow patterns related to the spontaneous breaking of the appropriate symmetry group \cite{siddharthamorse}. Once the coordinates of the $N$ points are found in this way, we need to simulate the behaviors of Carbon atoms arranged in that fashion, hence we essentially change the potential to the appropriate one, and perform a Density Functional type of computation. The number of atoms we can describe this way is of the order of $10^3$, highly demanding computer-time wise, but still too small for the reaching of the horizon. Nonetheless, the results obtained will be important to refine various details of the theory. We need to go further, towards a big radius of curvature $r$, when the Hawking effect should be visible.

Any serious attempt to understand Quantum Gravity has to start from the Hawking effect. That is why black holes are at the crossroad of many of the speculations about the physics at the Planck scale. From \cite{iorio} a goal that seems in sight is the realization of reliable set-ups, where graphene well reproduces the black-hole thermodynamics scenarios, with the analogue gravity of the appropriate kind to emerge from the description of graphene's membrane. The lattice structure, the possibility to move through energy regimes where discrete and continuum descriptions coexist, and the unique features of matter fields whose relativistic structure is induced by the spacetime itself, are all issues related to Quantum Gravity \cite{worldcrystal} that can be explored with graphene.

Many other tantalizing fundamental questions, can be addressed with graphene. To mention only two: There are results of \cite{mauriciosusy}, that point towards the use of graphene to have alternative realizations of Supersymmetry, and there are models of the Early Universe, based on (2+1)-dimensional gravity \cite{vanderbij}, where graphene might also play a role.

We are lucky that these ``wonders'' are predicted to be happening on a material that is, in its own right, enormously interesting for applications. Hence there is expertise worldwide on how to manage a variety of cases. Nonetheless, the standard agenda of a material scientist is of a different kind than testing fundamental laws of Nature. Therefore, let us conclude by invoking the necessity of a dedicated laboratory, where condensed matter, and other low energy systems, are experimentally studied with the primary goal of reproducing phenomena of the fundamental kind.

For the reasons outlined above, graphene is a very promising material for this purpose.

\section*{Acknowledgement}

The author warmly thanks the LISC laboratories of FBK and ECT*, Trento, Italy, for the kind hospitality, and acknowledges the Czech Science Foundation (GA\v{C}R), Contract No. 14-07983S, for support.

\end{document}